\def\year{2022}\relax
\documentclass[letterpaper]{article} 
\usepackage{aaai22}  
\usepackage{times}  
\usepackage{helvet}  
\usepackage{courier}  
\usepackage[hyphens]{url}  
\usepackage{graphicx} 
\usepackage{natbib}  
\usepackage{caption} 
\DeclareCaptionStyle{ruled}{labelfont=normalfont,labelsep=colon,strut=off} 
\usepackage{algorithm}
\usepackage{algorithmic}

\usepackage{svg}

\usepackage{newfloat}
\usepackage{listings}
\floatstyle{ruled}
\newfloat{listing}{tb}{lst}{}
\floatname{listing}{Listing}
\title{A Dataset of Coordinated Cryptocurrency-Related Social Media Campaigns}
\author{
    Karolis Zilius\textsuperscript{\rm 1},
    Tasos Spiliotopoulos\textsuperscript{\rm 2},
    Aad van Moorsel\textsuperscript{\rm 2}
}
\affiliations{
    \textsuperscript{\rm 1}Newcastle University, UK,\\
    \textsuperscript{\rm 2}University of Birmingham, UK\\
    kikarolis@gmail.com,
    a.spiliotopoulos@bham.ac.uk,
    a.vanmoorsel@bham.ac.uk
}

\usepackage{bibentry}

\begin{document}

\maketitle

\begin{abstract}
The rise in adoption of cryptoassets has brought many new and inexperienced investors in the cryptocurrency space. These investors can be disproportionally influenced by information they receive online, and particularly from social media. This paper presents a dataset of crypto-related bounty events and the users that participate in them. These events coordinate social media campaigns to create artificial "hype" around a crypto project in order to influence the price of its token. The dataset consists of information about 15.8K cross-media bounty events, 185K participants, 10M forum comments and 82M social media URLs collected from the Bounties(Altcoins) subforum of the BitcoinTalk online forum from May 2014 to December 2022. We describe the data collection and the data processing methods employed and we present a basic characterization of the dataset. Furthermore, we discuss potential research opportunities afforded by the dataset across many disciplines and we highlight potential novel insights into how the cryptocurrency industry operates and how it interacts with its audience.
\end{abstract}

\section{Introduction}
The last years have seen the rise of Decentralized Finance (DeFi), a form of finance that does not rely on central financial intermediaries, such as banks and exchanges, to offer traditional financial services, and instead utilizes smart contracts on distributed ledgers (blockchains). The cryptographically secured digital tokens that are stored on these blockchains, called cryptoassets, include cryptocurrencies (like Bitcoin and Ethereum), utility tokens, security tokens and Non-Fungible Tokens (NFTs) \cite{Karim2022}. A simple application of this is that people can effectively invest in a company or project by acquiring and trading digital tokens of that company without intermediaries. Because these cryptoassets are freely traded, there is a lot of room for price speculation and manipulation \cite{Nizzoli2020}. Importantly, compared to traditional investing, this decentralized investing can be conducted by lay investors \cite{Abramova2021} who are often influenced disproportionally by information they receive from social media \cite{Glenski2019,Jahani2018}. This provides a strong incentive for such crypto projects to create artificial "hype" in order to convince more social media users to buy their token, or in order to manipulate the price of a token in a certain way. In particular, they provide payments in the form of "bounties" to users willing to engage with the project on social media in a positive way, such as following, commenting, retweeting and creating content.

This paper presents a dataset of crypto-related bounty events that coordinate social media campaigns to create artificial hype around a crypto project, as well as the users that engage in these campaigns. Our dataset was crawled from the \textit{Bounties(Altcoins)} subforum of the \textit{Bitcointalk.org} online forum. While a lot of research has studied inauthentic, manipulative, deceptive and collusive content online, this type of bounty hunting phenomenon is unique due to the particularities of the cryptoassets context. In particular, the sentiment and the volume of information online have been found to strongly influence market prices and user investing decisions in traditional finance \cite{Bollen2011b}, and even more so in relation to cryptoassets \cite{Domingo2020,Xie2020}. In the context of cryptoassets, these coordinated campaigns aim to achieve this in two ways; directly, by influencing social media users to buy an asset based on artificial hype, and indirectly, by influencing trading algorithms that make investing decisions based on sentiment mined from social media \cite{Garcia2015}. What makes this practice particularly concerning is that bounty hunters are typically paid in the token that they are promoting. This gives them an incentive to "dump" (i.e., sell) the asset when the campaign ends, and since they have inside knowledge of exactly when the campaign will end, this practice can be considered equivalent to a "pump-and-dump" scheme \cite{Hamrick2021} that will dramatically crash the price of the token causing the investors outside the campaign to lose the funds they invested.

While this type of bounty hunting phenomenon is unique, it builds on known practices that have been identified and studied earlier in the literature. These practices have been tailored specifically to the crypto space and are of varying ethical and legal positioning and justification. \textbf{Bounty programs}, in general, have been commonly used to manage rewards traditionally aimed at incentivizing positive and productive behavior, such as finding bugs in software code \cite{Ding2019}, providing worthy answers in question-answer platforms \cite{Parnin2012} and verifying the bias and safety of AI systems \cite{Brundage2020}. The practices and ethics of \textbf{advertising}, in general \cite{Schauster2016}, and social media promotion \cite{Zeng2021}, in particular, have been debated by researchers. Related work has highlighted particularly questionable practices, such as \textbf{astroturfing} (i.e., hidden coordinated information campaigns that mimic genuine user behavior by incentivizing agents to spread information online) for consumers \cite{Kauppila2022, Lee2013b} and citizens \cite{Schoch2022}. \citet{Dutta2021b} have studied \textbf{blackmarket services} that are employed by "collusive" users to inflate the popularity of their online account and get appraisals for their content, while \textbf{spam} has been a topic of intensive research over the years \citep[e.g.,][]{Benevenuto2010,Yardi2009}. 

From the perspective of \textbf{finance}, researchers have examined the inner workings, the nuances and the relationships between cryptocurrency pump-and-dump schemes \cite{Hamrick2021}, market price manipulation \cite{Nizzoli2020}, and Initial Coin Offerings (ICOs) \cite{Ante2018}. From a \textbf{legal and regulatory} perspective, researchers are examining the ways and the extent to which social media content creators comply with the guidelines of the US Federal Trade Commission (FTC) which requires them to disclose their endorsements in order to prevent deception and harm to users \cite{Mathur2018}. Specifically in the crypto space, scholars are examining the practices of crypto users and businesses in order to determine the status of cryptoassets as \textit{securities} (i.e., investment contracts), something that would also place restrictions on discussions around cryptoassets and would require disclosure of any connections or endorsements \cite{Henderson2019}. Finally, such practices are in violation of the Terms of Service (ToS) of online platforms; for example Twitter's platform manipulation and spam policy\footnote{https://help.twitter.com/en/rules-and-policies/platform-manipulation} prohibits artificially amplifying content and makes specific mentions to coordinated activities for doing so. This mix of characteristics of other practices that is present in crypto-related social media bounty hunting muddles the way that campaign organisers, participants and even targets of the campaigns perceive the ethical and legal implications.

This paper presents a dataset collected from the \textit{Bounties(Altcoins)}\footnote{https://bitcointalk.org/index.php?board=238.0} subforum (message board) of the \textit{Bitcointalk.org}\footnote{https://bitcointalk.org/} forum describing crypto-related bounty events, the participants, the actions taken and the rewards received from 13-May-2014 when the first thread was started to 31-Dec-2022. We collected and analyzed information about 15.8K unique cross-media bounty events, 185K bounty hunters (i.e., users that participated in such events), 10.01M comments, 18M spreadsheet lines, 56K images, 42.1M Tweets and 33.9M links to Facebook posts, among other information posted on the subforum. We further enhanced the dataset to make it accessible to a broader set of researchers and enable future research in this area by linking information collected across the forum threads and aggregating it to calculate useful metadata. This resulted in a rich set of descriptive metadata; for example, we provide 31 data fields describing each Bounty event and 17 data fields describing each participant. 

While previous research has focused on detecting coordinated social media campaigns and attempting to infer the actors and the details of deceptive actions from their social media traces \citep[e.g.,][]{Peng2017,Keller2020}, our dataset provides comprehensive information on these campaigns and their organisation from their source, including the structure of incentives, exact instructions and timing of actions, and a full account of all participants. We expect that this unprecedented detail afforded by our dataset will be invaluable to researchers studying online coordinated deceptive practices.

We make the entire dataset, along with a smaller sample and supplementary material, publicly available at the following link: (Dataset URL: \url{https://zenodo.org/record/7539178}).

\section{The Bitcointalk.org Forum and the Bounties(Altcoins) Subforum}
The BitcoinTalk forum was created by the developer of Bitcoin in 2009, the same year that Bitcoin was developed\footnote{https://bitcointalk.org/index.php?topic=5}. While originally focusing on Bitcoin itself, the discussions on the forum gradually expanded to include alternative cryptocurrencies (altcoins). The forum currently has more than 3.5M registered users and 61.5M posts on 1.3M topics\footnote{https://bitcointalk.org/index.php?action=stats}.
As the first and very popular community created around cryptocurrencies, the forum had an original intention of furthering adoption and supporting users. Areas of discussion include technical aspects of distributed ledgers, such as development and mining, economics, politics, as well as supporting users in using the technology, trading, and recovering from scams. As cryptocurrencies became more widespread and less technically-inclined users turned to the forum for information and advice, researchers utilized the BitcoinTalk forum to study crypto-related collective intelligence \cite{Jahani2018}, speculation \cite{Xie2020} and scams \cite{Vasek2019}.

The main BitcoinTalk forum comprises multiple subforums, each focusing on a separate aspect of cryptoassets. Each subforum contains many discussion threads initiated by different users. The earliest entries of the Bounties(Altcoins) subforum date back to 2014, although it didn't receive much attention until 2017. The subforum experienced a massive growth in popularity during 2017 – 2018, which aligns with the first steep increase in crypto adoption. More than 15,000 threads have been created since and it has evolved to the most popular subforum of BitcoinTalk\footnote{see note 5}.

Bounties(Altcoins) serves as a place to organize airdrops (cryptocurrency giveaways) and alert the public about upcoming projects or ICOs, however it is dominated by \textit{bounty events}. A bounty event involves the distribution of cryptocurrency or monetary rewards for the accomplishment of event-specific objectives. These rewards are typically distributed in the token being promoted, although it is not uncommon to use stablecoins (e.g., USDT) and even US dollars transferred via a service like Paypal. Each bounty event can have one or more \textit{campaigns} which are allocated a percentage of the total reward pool. The vast majority of the campaigns require participants to use their social media accounts to spread awareness about the project, however event organizers may also be interested in producing quality content with campaigns such as translation (e.g., of a whitepaper or  website content), copywriting, or visual content creation such as infographics or visual art.

Over time, certain norms have arisen to make it easier for users to participate in a bounty event. Although for some events the procedure differs slightly, most events follow a three-part structure:
\begin{enumerate}
\item A user shows their interest in participating by posting a “proof of registration” comment and including information about themselves (e.g., Telegram username, campaigns joined, crypto wallet address) which may differ slightly from event to event.
\item The participant completes the tasks defined in the campaign rules section. The event can last from a few weeks to a few months or even up to a year, therefore the tasks can be completed more than once (usually once a week).
\item The participant provides evidence of engagement with the project by posting “proof of participation” comments that include some required information (e.g., social media handle, social media interaction links) which is usually similar for all events. However, a small subset of events uses other means for providing evidence, such as Google forms.
\end{enumerate}

Users that employ social media bot accounts are  identified by the event organisers and are removed from the event. Participants are rewarded for their contribution at the end of an event. Each campaign has different amounts of stakes that are distributed to users which is proportional to the difficulty of the task, the quality of content, or the value provided by the user (e.g., a user with more Twitter followers will be allocated more stakes for the same task). The total reward can be calculated with the  formula
\[User \: Rewards= \frac{Campaign \: Prize \: Pool}{Distributed \: Stakes} * User \: Stakes\]

\section{Data Collection}
In this section we describe our data-scraping methods. The process consisted of crawling \textit{subforum} pages, \textit{user} pages, \textit{thread} pages, \textit{images}, and \textit{Google spreadsheets}. To achieve this, a Python crawler was developed which used HTML requests library\footnote{https://github.com/psf/requests} to fetch images and information from forum pages, BeautifulSoup\footnote{https://github.com/waylan/beautifulsoup} to parse HTML and gspread library\footnote{https://docs.gspread.org/en/v5.7.0/} to access Google Sheets API endpoints.

\subsubsection{Subforum and thread crawling}
The forum comprises multiple subforums, which contain a collection of threads divided into pages. Each subforum has a unique identifier; Bounties(Altcoins) ID is 238. Each page in the subforum contains up to 40 threads and has a unique identifier which starts at 0 for the first page and is increased by 40 for every subsequent page. Additionally, each subforum page contains a navigation strip, which can be utilized to retrieve the last page ID. This allowed us to generate a unique URL for each page by inserting the identifiers into \textit{https://bitcointalk.org/index.php?board=238.page\_id}. Crawling all pages resulted in retrieval of information from 15K unique events.

\subsubsection{Comments crawling}
Each forum thread has a unique identifier retrieved during subforum crawling. Similarly to subforum page, each thread has a comment page ID which starts at 0 for the first page and is increased by 20 for each subsequent page and a navigation section which can be used to derive the last comment page ID. This allowed us to generate unique URLs for each comment page by inserting unique identifiers into \textit{https://bitcointalk.org/index.php?topic=thread\_id.page\_id}. 10M comments were obtained after crawling each page.

\subsubsection{User profile crawling}
Forum users are given a unique identification number which was retrieved during comment crawling. Inserting it into \textit{ https://bitcointalk.org/index.php?action=profile;u=user\_id} allowed us to access each user’s profile page. Information on 185K users was retrieved after crawling every page.

\subsubsection{Image crawling}
Some events in the forum used images to convey information, which had to be retrieved, converted to text and inserted back into the main post for more accurate processing. The image URLs could be found in HTML \verb|<img>| elements of the main forum post. In total, 64K unique image URLs were detected out of which 56K were downloaded successfully and 8K were no longer being hosted.

\subsubsection{Spreadsheet crawling}
Some events used Google spreadsheets to track user progress and keep user information in one place. Spreadsheet URLs were collected from main posts using regular expressions to match spreadsheets domain\footnote{https:///www.docs.google.com/spreadsheets} and cleaned to extract unique spreadsheet identifiers. Using this information, all rows were retrieved from Google sheets by querying Google Sheets API with gsread. This resulted in the retrieval of more than 20M distinct rows from Google spreadsheets.

\section{Data Processing}
In this section we provide an overview of how the collected data were processed and labeled and we explain the structure of the dataset.

\subsection{Forum Threads}
The primary focus of the dataset is bounty events. This section explains the methods used to extract information from the main post as well as how such events were distinguished from other sorts of events in the subforum.

\subsubsection{Categorization}
Information from 15,870 unique events was fetched from the first thread creation on 13-May-2014 to 31-Dec-2022. Each thread was assigned one or more categories based on the title, main post comment and user replies. The category of \textit{Bounty} was assigned to threads that:
\begin{enumerate}
\item Had seven or more out of 14 common Bounty event traits. Nine of the traits involved matching against a single word (Stake, Rewards, Campaign, General rules, Token, Bounty, Pool, Spreadsheet, Whitepaper). Five traits had multiple synonyms (Rules: 3, Reward allocation: 19, Proof: 2, Signature: 2 and Social media campaigns: 22); or
\item Had six out of the 14 common bounty event terms and more than 100 replies or the terms included \textit{Bounty} and \textit{Whitepaper}; or
\item More than 50\% of comments were classified as registration / participation (see next section for details) and the event was closed (the main post, title or both removed) or had more than 200 replies.
\end{enumerate}

Events containing the keyword \textit{bounty} in the title but not meeting any of the requirements above were assigned a category of \textit{Bounty(low quality)}. These categories had to be separated because titles were created by users and some threads that contained the \textit{bounty} keyword in the title had nothing to do with bounty hunting.

\begin{figure*}[t]
  \centering
  \includegraphics[width=\textwidth]{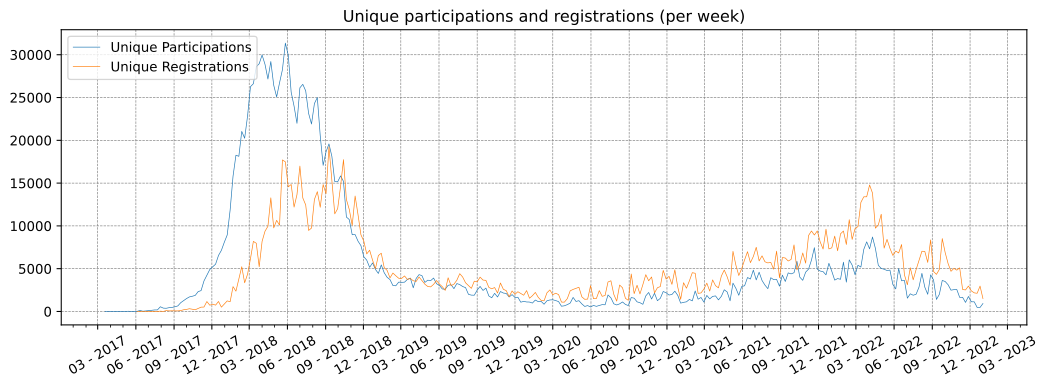}
  \caption{Number of unique participations and registrations per week (multiple comments of the same type by an individual in the same event are counted as one).}
  \label{fig:RSUencountered}
\end{figure*}
Other categories included: \textit{Moved} (event moved to another subforum), \textit{ICO} (Initial Coin Offering), \textit{Closed} (removed content), \textit{Announcement}, and \textit{Other}. 7,177 events were identified as Bounty events, 3,432 threads were identified as Bounty (Low Quality) and 5,261 events were assigned one of the other categories. While just fewer than half of the threads were categorized as Bounty events, 96.4\% of the collected comments were posted in Bounty events and 1.2\% to Bounty(Low Quality) events, demonstrating the importance of Bounty events in the subforum.
\subsubsection{Content processing}
Information about forum events was gathered from several places. Data which are included in every thread were collected from the subforum page which is well structured. This information includes \textit{thread title}, \textit{thread author}, \textit{username} and \textit{unique ID}, number of \textit{replies} and \textit{views}, and information about last comment: \textit{date}, \textit{author username} and \textit{ID}. The thread page also contained structured information that was common between all events such as event creation date and main post in HTML format. Further analysis was only carried out on events assigned a category of Bounty.

All HTML \verb|<a>| elements of the main post were inspected for links. By matching domain names with regular expressions, URLs were divided into the following categories: \textit{spreadsheet}, \textit{image} – embedded images that were converted to text, \textit{forum} – bitcointalk.org domain, \textit{social media} and \textit{other} – the rest of the URLs.

The remaining event information was obtained from text. To overcome a BeautifulSoup parser limitation when converting HTML to text and to preserve post structure, a new line character was inserted after each element that was not \verb|<br>| or inline. The text was split on new lines to aid the analysis process.

The \textit{Reward pool} size section of the event was identified by matching keywords (e.g., "Bounty pool", "Bounty details"). Subsequent lines were checked for three common reward pool patterns (e.g., "\$100 worth of ABC", "ABC 500", "200\$ in ABC token" ). 

\textit{Token names} were detected in title, reward pool and text (near phrases such as "Token" or "Coin"). Each of these strings was tested against a set of conditions: two characters or longer, no words in a list of 27 phrases (e.g., “NFT”, “USDT”, “APY”), only contain characters A-Z and \$, not a word in a USA or UK dictionary. Strings that passed all checks and were most frequent are included in the dataset.

\textit{Reward allocation} was determined by matching phrases (e.g., "Reward allocation", "Budget"). Each subsequent line was inspected for \verb|%| sign and campaign name. If the percentages of one or more campaigns added up to a 100\%, then a list of campaign titles and reward allocation percentages was recorded in the database.

Bounty events would often have sections for event \textit{general rules}, and  \textit{rules} and \textit{rewards} for each campaign. Retrieving these sections comprised two steps; detecting in which line the section began and where it ended:
\begin{enumerate}
\item The general rules section was detected by matching the text for relevant phrases (e.g., "Bounty rules"). For each campaign this was achieved by matching a phrase (e.g., "Twitter campaign") followed by the keywords “Rules” and “Rewards”.
\item Once the index of a line where the section began was found, the subsequent lines were checked for an ordered or unordered list. If found, this information outlined where the section began and finished.
\end{enumerate}
\subsection{Thread Comments}
\subsubsection{Categorization}
The dataset contains five comment types: \textit{mixed} (participation and registration in one), \textit{participation}, \textit{registration}, \textit{author} (comments by the thread author) and \textit{other}. Comment category was assigned with a 4-step process:
\begin{enumerate}
\item Check if userid of comment author and main post author match. If yes, assign author category, else next step.
\item Check how many out of six common proof of registration traits (headline “\#proof of authentication”, forum username, forum profile URL, telegram username, campaigns joined, wallet address) the comment has.
\item Check how many out of six common proof of participation traits (headline “week” or “day”, social media URLs, social media campaign names, social media username, numbered list, keywords such as "Like" and "Retweet") the comment has.
\item If a comment passes a registration threshold of 4 traits (3 if there are fewer than 5 lines in the comment) or has \# symbol with one of 75 synonyms for “proof of authentication” then the threshold for participation comment becomes 2 (it must contain social media links). If both thresholds are passed, the comment is assigned a category of mixed else, it becomes registration. If registration threshold is not passed, the comment is considered as participation with threshold of 3 (must contain social media links or keywords: “week”, “day”). If this threshold is passed the comment is assigned a category of participation, else it is assigned to other.
\end{enumerate}

One limitation of this process is detecting registration comments which include little information, however it is very efficient at identifying and distinguishing between registration and participation comments. We report the results of a manual check of a random sample of 100 comments of each category for different threshold values in the Supplementary Material. We also provide the regular expressions used to match social media URLs. 

\subsubsection{Content processing}
Each comment shares some mutual information retrieved from the forum: comment id, author username and id, post time. Additionally, the thread id where the comment was found is included in the dataset.
Comments of type \textit{other} and \textit{author} were not processed and contain the comment text saved in HTML format. This choice was made to preserve the structure of the comment since it would be lost if the comment was a reply (i.e., contained one or more \verb|<div class="quote">| elements).

On the other hand, \textit{participation}, \textit{registration} and \textit{mixed} comments were processed extensively. Mixed comments include a Boolean check and get analyzed by both algorithms. Most of the information was extracted with conditional statements and regular expressions utilizing library re\footnote{https://docs.python.org/3/library/re.html}.

Registration comments include information such as \textit{campaigns a user registered for}, \textit{Telegram username}, \textit{Twitter username}, \textit{forum rank} and \textit{post count} (specified by user), \textit{line number} in spreadsheets and \textit{crypto wallet address}. Additionally, the dataset also contains \textit{other information} which was rare and did not get assigned any label.

Participation comments include information about user \textit{social media username} (with social media platform), \textit{campaigns participated} in (based on participation links found) and all proof of interactions which include \textit{TweetIDs} and social media links (\textit{Twitter (without id)}, \textit{Facebook}, \textit{Instagram}, \textit{Telegram}, \textit{Reddit}, \textit{YouTube}, \textit{Medium}, \textit{LinkedIn}, \textit{Discord}, \textit{TikTok}, \textit{Steemit}, \textit{Image sharing}, \textit{Other})
\subsubsection{Google spreadsheets}
Each Google spreadsheet contains one or more sheets, which are processed independently. The process comprises two parts:
\begin{enumerate}
\item Many spreadsheets included banners of varying height, so we had to detect which row contains column names by iterating through the first 15 rows and checking which one has the most out of ten common column names (e.g., "timestamp", "twitter followers", "post count").
\item Label consequent rows based on column names.
\end{enumerate}

Unidentified columns \textit{labels} and \textit{data} are also included in the dataset. Identified columns include: \textit{timestamp}, \textit{proof of registration post URL}, \textit{forum rank}, \textit{forum profile URL}, \textit{forum username}, \textit{wallet address}, \textit{email address}, \textit{stakes}, \textit{twitter followers}, \textit{twitter audit URL}, \textit{post count}, \textit{social media username}, \textit{social media profile URL}.

\subsection{Images to Text}
Using Tesseract OCR\footnote{https://github.com/tesseract-ocr/tesserac}, locally stored images were converted to text and inserted into respective HTML \verb|<img>| elements. The process encountered a few limitations when dealing with intricate graphics pictures. Occasionally when dealing with such images, there were misinterpreted letters, random symbols, or parts of the text missing. The dataset contains the \textit{text} generated from locally retrieved images and corresponding \textit{image URLs}.

\begin{table}[t!]
\begin{center}
\begin{tabular}{ |l|l| }
 \hline
Data category &	Entries collected \\
 \hline
Total threads in subforum	& 15,870 \\
Bounty events	& 7,177 \\
Total comments in subforum & 10,024,001 \\
Total comments in Bounty events & 9,655,442 \\
Bounty comments (participation)	& 6,926,577 \\
Bounty comments (registration)	& 1,670,871 \\
Bounty comments (mixed) & 233,268 \\
Bounty comments (author) & 107,650 \\
Bounty comments (other)	& 1,183,612 \\
Social media links & 82,854,734 \\
Images (converted to text)	& 56,529 \\
User profiles	& 185,709 \\
Google spreadsheets  & 24,887 \\
Lines in Google spreadsheets	& 20,182,746 \\
Labeled Google spreadsheets entries	& 18,082,209 \\

 \hline
\end{tabular}
\end{center}
\label{demo-table}
\caption{Statistics of collected dataset.}
\end{table}

\subsection{Structure of the Processed Dataset} Our data collection and analysis process resulted in a PostgreSQL database with twelve tables. We exported the tables to twelve tsv files and added some additional annotations. The final dataset contains four tsv files of unprocessed information retrieved from the subforum, google spreadsheets, and image hosting services. To accommodate the largest audience possible and to aid further analysis by researchers, we also release eight processed tsv files resulting from the cleaning, merging and annotation conducted in the processing stage. In total, the files contain 159 fields (events – 31 fields, users – 17, comments\_participation - 29, comments registration – 21, comments\_author, comments\_other and comments\_raw – 7, rewards\_and\_rules – 4, spreadsheets – 18, spreadsheets\_raw – 5, images\_raw – 3, threads\_raw – 10). More extensive description of each file and field can be found in the Supplementary Material that come with the dataset. Table 1 shows some statistics of the collected data.

\section{Exploratory Analysis}
\subsubsection{Subforum activity over time}
Figure 1 shows forum activity in terms of \textit{unique} user registrations and participations. If a user created more than one participation/registration comments in the same thread, they were only counted as one participation/registration. To preserve plot space, the visualization starts from 2017 as there are only ten threads that were created before that time. From the illustration we can observe that bounty events first started gaining traction in the second half of 2017 and peaked during the first half of 2018, which aligns with the timeline of the first massive bull market run\footnote{https://coinmarketcap.com/charts/} and corresponding increase in crypto adoption. During this period we can observe up to 30,000 unique participations every week. Since then, the forum has stayed active with some ups and downs in participation over time. After a substantial slump in 2019 and 2020 where activity fell to almost one tenth of the peak, activity started to slowly recover in 2021 and falling again from 2022.

\subsubsection{Size and activity of Bounty events}
Most Bounty events are substantially large to make a difference in promoting a project; 4,402 (61.4\%) events had more than 100 unique participants and 837 (11.7\%) had more than 1,000 unique participants. We found five outlier events where more than 5,000 forum users participated (Figure 2).

The number of activities that participants engaged in varied substantially across Bounty events. In the early days of the subforum the norms hadn't evolved yet and the comments did not have the structure that could help us crawl them successfully and categorize them as registration or participation comments. Still, a substantial number of events show relatively little activity with limited participation from users. On the other hand, we also identified a large number of events with several thousand participation comments; each one typically being a weekly comment including many activities across different campaigns (Figure 3).

\begin{figure}[t!]
  \begin{center}
  \includegraphics[width=\linewidth]{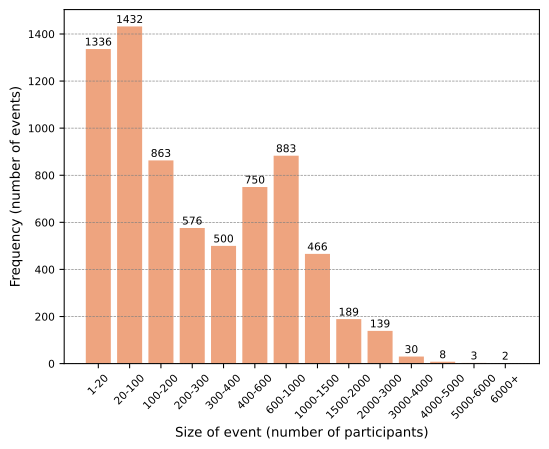}
  \end{center}
  \caption{Number of unique participants per Bounty event. Note that for visualization purposes, the bins in the horizontal axis are not equal.}
  \label{fig:RSUencountered}
\end{figure}

\begin{figure}[t!]
  \begin{center}
  \includegraphics[width=\linewidth]{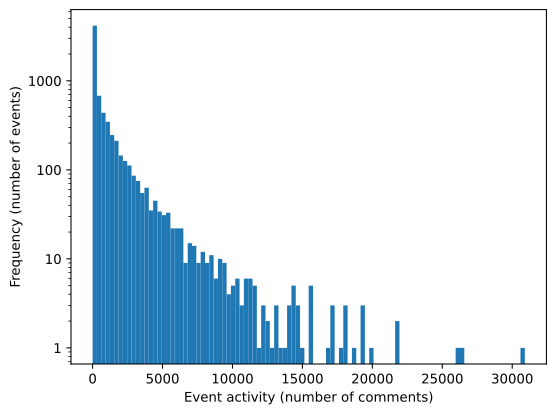}
  \end{center}
  \caption{Number of participation comments per Bounty event. Note that the y axis is logarithmic.}
  \label{fig:RSUencountered}
\end{figure}

\subsubsection{Duration of events}
Figure 4 shows the duration of Bounty events. The duration is determined by calculating the time passed from the first post in the thread (which announces the beginning of the event) to the last (which typically ends the event). An event typically requires participants to post every week. However, sometimes, the event can get paused and the thread blocked for a few days. For this reason, we considered any period of more than three weeks between two consecutive comments as the end of an event.

We can observe that 350 (5\%) events last less than a day and 30\% end in the first month, which leads us to believe that the majority of Bounty events end successfully. A small number of events (92) kept operating for more than 300 days. These projects may have a long-term objective to remain in the cryptocurrency market and invest in a continuous bounty program.

\begin{figure}[t!]
  \begin{center}
  \includegraphics[width=\linewidth]{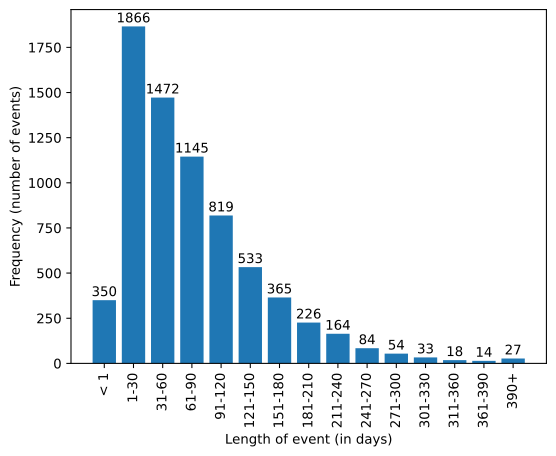}
  \end{center}
  \caption{Number of consecutive days a Bounty event stayed active.}
  \label{fig:RSUencountered}
\end{figure}

\subsubsection{Bounty campaign popularity}
We discuss which campaigns are most likely to attract participating users and how many links grouped by campaign type have been collected from user comments.

Figure 5 illustrates how many unique participations for each campaign type were recorded in the subforum each week. Twitter and Facebook are the most utilized platforms across all time by a substantial margin. Telegram and Instagram started being utilized considerably in recent years.

In Figure 6 we can observe campaign popularity from another angle. Twitter and Facebook are still the most popular platforms with 42.1M and 33.9M URLs collected respectively, while all other platforms account for less than 9\% of all links collected. No Telegram links were collected as these tasks did not require users to share a proof of participation link. Campaigns that involved more complex or time-consuming tasks, such as creating YouTube videos or Medium blog posts are underrepresented compared to their relative importance.

\subsubsection{Reward allocation}
Bounty events typically involve more than one campaign, necessitating the distribution of the reward pool across campaigns. Each campaign gets allocated a percentage of the total pool based on the project owners' preferences and the goal of the event. We were able to determine reward allocation that added up to 100\% for only 2,705 of the 7,170 Bounty events, which we analyze below. Figure 7 shows the average reward allocation for each campaign type as a percentage of the full available reward pool. Campaigns that were encountered fewer than five times are grouped under the category “rare campaigns”.

The campaign with the most funds allocated on average is \textit{Signature}. Signature is an exclusive campaign to the forum that requires users to update their profile with a specific banner that is visible on all their posts, effectively acting as advertisements for the Bounty event. This suggests that event organizers may expect that forum users will consider campaigns with signatures to be more professional, trustworthy or popular, and thus will be more likely to join them.

The second most rewarded campaign type is \textit{Content creation}. This campaign type includes activities such as .gif and meme creation, infographics, and copywriting. Together with the \textit{Rare} category, which also attracts a substantial level of rewards, this suggests that Bounty events are particularly interested in unconventional, but also sophisticated and time-consuming activities of project promotion.

Twitter and Facebook are the third and fourth most rewarded campaigns, which is expected since they make up 91\% of all links collected from proof of participation comments and are key channels for spreading information about Bounty events online.

\begin{figure*}[t]
  \centering
  \includegraphics[width=\textwidth]{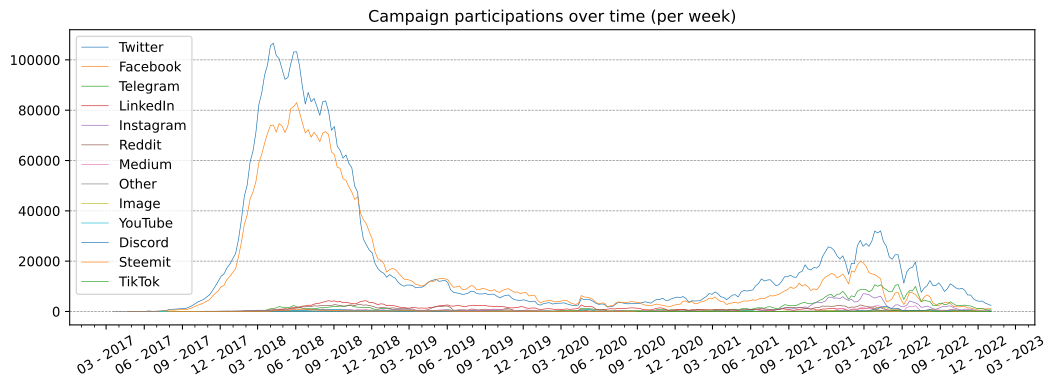}
  \caption{Number of unique participations (multiple participation comments from the same user in the same thread are counted as one) for different campaign types over time. The legend is sorted from most to least popular campaigns, top to bottom.}
  \label{fig:RSUencountered}
\end{figure*}

\begin{figure}[t!]
  \begin{center}
  \includegraphics[width=\linewidth]{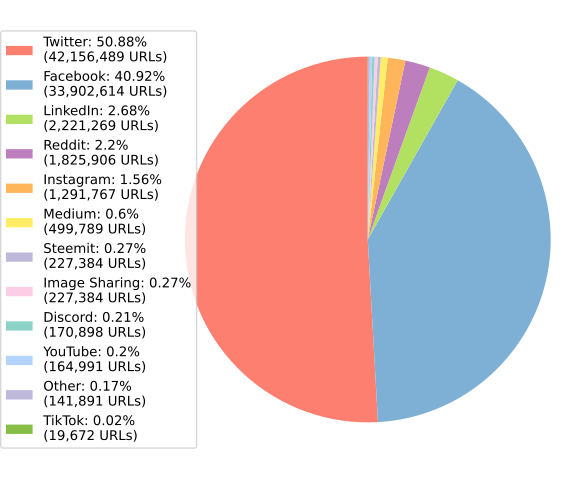}
  \end{center}
  \caption{Number of links collected from proof of participation comments for different types of campaigns.}
  \label{fig:RSUencountered}
\end{figure}

\begin{figure}[t!]
  \begin{center}
  \includegraphics[width=\linewidth]{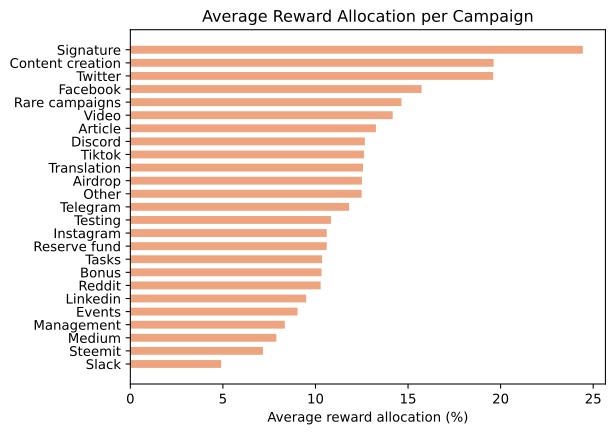}
  \end{center}
  \caption{Average percentage of reward pool that gets allocated to different campaign types in Bounty events.}
  \label{fig:RSUencountered}
\end{figure}

\section{Research Opportunities Using the Dataset}
The BitcoinTalk forum has a long history and has had significant influence in the cryptocurrency space. As the largest subforum, and one that encourages users to engage with social media, create content and participate in discussions, the Bounties (Altcoins) dataset can provide useful insights to researchers of various interests. Here we discuss some unique research opportunities arising from our dataset.

\subsubsection{Financial analysis} Market price manipulation has long been of interest to researchers, both in the stock market and in the crypto space \citep[e.g.,][]{Gandal2021}. Further studies can use our dataset to study the effect of coordinated social media campaigns \cite{Schou2022} by examining the price movement of the tokens included in our dataset. This can show whether the bounty campaigns have been successful and whether they were intended to con investors out of their investments by inflating the price and then "dumping" the token \cite{Mirtaheri2021,Nizzoli2020}. Further research can also attempt to correlate the activity in the forum or the prices of the promoted tokens with the movements of the wider economy, Bitcoin or other cryptoassets \cite{Gandal2021} to examine how these affect online campaigns.

\subsubsection{Social media and network analysis} A lot of research has examined how and why people use multiple media \cite{Spiliotopoulos2020a, Alhabash2017}. Our dataset, which includes the details of multiple social media accounts for each user, can be valuable for social media researchers studying cross-media user posting practices. Researchers can also follow the social media links from the dataset to gain a better understanding of the effect of these campaigns. For example, how much of the discussion around a crypto project is organic and how much is driven by bounties? How many followers of a crypto project are genuine and how many are bounty hunters? Social network researchers can use the bounty hunter profiles provided in the dataset to determine details about the networks \cite{Himelboim2017} of bounty hunters; How many are connected and follow each other within and across media? What cliques are they part of? Are their followers genuine or are they just following one another and effectively promoting the project to the converted?

\subsubsection{Natural Language Processing} Researchers can follow the social media links and download text content from the Tweets, Facebook posts, Reddit discussions and blog posts to study the specific language used by bounty hunters. Natural Language Processing (NLP) techniques \cite{Hirschberg2015} can identify the topics discussed and the linguistic characteristics of this content, in order to detect it and protect social media users. Sentiment analysis techniques \cite{Valle-Cruz2022} can be used to quantify the effect of changes in online sentiment created by bounty campaigns on the crypto market \cite{Xie2020}.

\subsubsection{Policy and regulation} Legal scholars and policy makers can use the dataset to reach a more nuanced understanding of the crypto space and the relationship between social media or online communities and cryptocurrency investing. Examination of the structure and organization of Bounty events and the practices of crypto bounty hunters will inform the evaluation of the current regulatory state and protect naive investors. This is particularly important as the calls for regulation of the crypto space become increasingly insistent \cite{Henderson2019}.
\section{Broader Impact of the Work and Ethical Considerations}
Overall, there are numerous benefits that can be derived from our dataset. We bring awareness to a large number of Bounty events and campaigns that can potentially lead inexperienced crypto investors to make poor financial decisions. Our dataset can be a useful resource for researchers that can study the details and structure of the campaigns, gain insights into how the cryptocurrency industry operates and how it interacts with its audience, and develop ways to protect users and inform policy makers.

Of course, these benefits need to be considered together with the potential risks of collecting the data and sharing the dataset, especially risks related to unanticipated secondary use \cite{Salganik2019}. With regards to the data collection stage, while our dataset includes information and links from a range of social media and other online platforms, we only collected data from the BitcoinTalk forum. This data collection is in line with the Terms and Conditions of the forum, and, in fact, the forum specifically encourages data scraping from its boards\footnote{https://bitcointalk.org/index.php?topic=5208423.0}. The collected raw data were held in secure password-protected devices and cloud accounts.

A cross-media user dataset presents additional potential misuses compared to a dataset from a single social media platform, such as more extensive profiling and tracking, cyberstalking, and identity theft. In order to minimize the risks associated with sharing the dataset, we removed email addresses. However, we decided to keep other account information, such as social media account IDs and crypto wallet addresses. It was clear to us that these accounts were created for the purpose of participating in Bounty events and there is minimal overlap with forum users' personal social media accounts or other personal information. The removal of personal information and the use of non-personal (or "throwaway") social media accounts by the forum users keeps the risks associated with sharing the dataset relatively low. It is possible that a forum user in the future may decide to delete their account from the forum and their account information will remain in our dataset, but we consider that to be in line with the reasonable privacy expectations of public forum users, especially since a significant amount of content scraping takes place by other forum users in public\footnote{see previous note}. We do not foresee that the release of our dataset can put any of the forum users in any additional danger or risk, as the activities taking place during these Bounty events are currently within the law. Furthermore, while we did not employ any social media APIs to collect the data and, thus, we are not technically subject to any of their Terms and Conditions, we still decided to not share social media content (e.g., tweets) and to only share social media URLs and Tweet IDs. Finally, this work received ethical approval from our institution.

\subsubsection{Compliance with FAIR principles}
Our dataset adheres to the four FAIR data principles\footnote{https://force11.org/info/the-fair-data-principles/}. Our dataset is \textit{Findable} as it is assigned a unique and persistent Digital Object Identifier (DOI): \verb|10.5281/zenodo.7539178|. The dataset is indexed in a searchable resource and contains rich metadata that describe it. The data and metadata are retrievable by their identifier in an open and free manner, making our dataset \textit{Accessible}. The dataset is released in tsv format, making it \textit{Interoperable}. It is extensively documented with metadata describing every data field in a text file that accompanies the data. We also provide further details of the regular expressions used for some of the data processing we conducted in a supplementary file and we also share files with the raw collected data in order to enable unanticipated future uses. This makes our dataset \textit{Re-usable}.

\section*{Acknowledgments}
The reported research is supported by UKRI under Grant title: AGENCY: Assuring Citizen Agency in a World with Complex Online Harms. Grant reference: EP/W032481/2.
Further support was received by the EPSRC Centre for Digital Citizens. Grant reference: EP/T022582/1.

\bibliography{refs}

\end{document}